\documentclass[prl,twocolumn]{revtex4}
\usepackage[dvips]{graphicx}%
\usepackage{amsmath}
\usepackage{bm}
\begin{document}
%%%%%%%%
%  \small
%%%%%%%%
\title{Non-Gaussian Entangled States}
%\author{Ryo Namiki}% / \today
%\email[Electric address: ]{namiki@qi.mp.es.osaka-u.ac.jp}%{namiki@qo.phys.gakushuin.ac.jp} 
%
%\author{Tetsushi Takano}\author{Yoshiro Takahashi} %\email{takano@scphys.kyoto-u.ac.jp}
%\affiliation{Department of Physics, Graduate School of Science, Kyoto University, Kyoto 606-8502, Japan}
%\author{Shin-Ichi-Ro Tanaka}
%\affiliation{Department of Physics, Graduate School of Science, Kyoto University, Kyoto 606-8502, Japan}
\author{Ryo Namiki}\affiliation{Department of Physics, Graduate School of Science, Kyoto University, Kyoto 606-8502, Japan}\email[Electric address: ]{namiki@qi.mp.es.osaka-u.ac.jp}
%\author{Yoshiro Takahashi}\affiliation{Department of Physics, Graduate School of Science, Kyoto University, Kyoto 606-8502, Japan}\affiliation{CREST, JST, 4-1-8 Honcho Kawaguchi, Saitama 332-0012, Japan}

%\author{Masato Koashi}\author{Nobuyuki Imoto}\affiliation{CREST Research Team for Photonic Quantum Information, Division of Materials Physics, Department of Materials Engineering Science, Graduate school of Engineering Science, Osaka University, Toyonaka, Osaka 560-8531, Japan}
\date{\today}%March 17, 2007}%
\begin{abstract} 
We provide a family of Non-Gaussian pure entangled states in a bipartite system as the eigen states of a quadratic Hamiltonian composed of the Einstein-Podolsky-and-Rosen-like operators. The ground state of the Hamiltonian corresponds to the two-mode-squeezed vacuum state, and belongs to the Gaussian states. In contrast, all of the excited states are Non-Gaussian states. A separable inequality maximally violated by the eigen states 
is derived. % 
\end{abstract}

%\pacs{03.67.Dd, 42.50.Lc} 
%\keywords{quantum cryptography}
\maketitle

%\newpage
%%%%%%%%%%%%%%%%%%%%%   part 1 %%%%%%%%%%%%%%%%%%%%%%%%%%%%%%%%%%%%%%%%%%%%%%%%
%%%%%%%%%%%%%%%%%%

The number states are fundamental states to describe quantum nature of the world %\cite{NS} \cite{GCS}
 whereas the coherent states are more accessible Gaussian states in quantum optics and of relevance to the Gaussian description of the nature. The single-photon state and more-than-one-photon number states are typical example of Non-Gaussian (NG) states and of essential for the quantum information processing (QIP) \cite{NC00,RMP-CV}. In the combined systems, the entanglement property of the states are one of the central objective on the modern physics as well as on QIP \cite{EPR35,NPT}. For the class of Gaussian states, the characterization of entanglement has been developed \cite{Duan-Simon, Simon, RMP-CV, Gie03, AdessoRev}, and there are several approaches to take into account the NG states \cite{SV1,Dell06,SV2,NGsep,AdessoRev}. However, it seems that the number of the examples of the NG entangled states is rather limited and the testing ground for entanglement theories concerning the NG states is lacking. %NG-Entangle-2,NG-Entangle-3,

In this Letter, we introduce a family of NG pure entangled states as the eigen states of a Hamiltonian that describes two harmonic oscillators. % whose ground state is the two-mode-squeezed vacuum (TMSV) state. 
The states are considered to be a typical example of the NG states similar to the case of the number states in the one mode system, and serve as a useful tool to analyze the entanglement in infinite dimensional systems beyond the Gaussian approach.

In the seminal paper by Einstein, Podolsky, and Rosen (EPR) \cite{EPR35}, 
 the nonlocal property of quantum continuous variable system was demonstrated by introducing simultaneous eigenstates of the EPR operators $\hat x_A - \hat x_B$ and $\hat p_A+\hat p_B$, % Theethose are uncommutable on each of subsystems $$, but commutable on the rotal system, $[x_A-x_B, p_A+p_B] = 0$. 
those are uncommutable on each of the subsystems
%\begin{eqnarray}
$[\hat x_A, \hat  p_A]=[ \hat x_B, \hat  p_B]=i$, % \end{eqnarray}
but commutable on the total system, %\begin{eqnarray}
$[ \hat x_A-\hat x_B, \hat p_A+ \hat p_B] = 0$. % \end{eqnarray} 
The essential point for nonlocality is that the total uncertainty is smaller than the sum of the subsystem's uncertainty, and such property can be seen % reproduced 
for the EPR-like operators %$\frac{ x_A }{\sqrt \xi} - \sqrt \xi  {x_B }=:  \omega_+(\xi)$ and $   \frac{ p_A }{\sqrt \xi}  +  \sqrt \xi   {p_B} =:\omega_- (\xi) $. 
\begin{eqnarray}
 \hat \omega_+(\xi) &:=&\frac{ \hat x_A }{\sqrt \xi} - \sqrt \xi \hat  x_B \nonumber \\ 
 \hat\omega_- (\xi)& := &\frac{ \hat p_A }{\sqrt \xi}  +  \sqrt \xi   {\hat p_B}.
 \end{eqnarray}%whose 
% which correspond to the EPR operators in $\xi \to 1$. %are 
Associated with this pair of the EPR-like operators, Duan \textit{et al.}, provided the separable criterion \cite{Duan-Simon}: Any separable state must satisfy % 
\begin{eqnarray}
\langle\hat\Omega  (\xi) \rangle &\ge& \xi^{-1}+ \xi.  \label{Duan-c} \end{eqnarray}
 Here the total noise of the EPR-like operators is defined by   
\begin{eqnarray}
 \hat\Omega  (\xi) &:=&  \Delta^2 ( \hat \omega_+(\xi) )+ \Delta^2 ( \hat \omega_-(\xi) )  \label{noise-op}%
\end{eqnarray} and $\Delta \hat O := \hat O -\langle \hat O \rangle $ represents the deviation of the operator.
%Here the variance is defined by  
In what follows we assume $\langle \hat  x_A \rangle =\langle  \hat x_B \rangle =\langle  \hat p_A \rangle =\langle  \hat p_B \rangle =0 $ and $0 < \xi<1 $. %the condition on the parameter $0 < \xi<1 $. % without loss of generallity.

From the commutation relation $[ \hat\omega_+(\xi),  \hat \omega_-(\xi)]= i\left( \xi^{-1}-\xi \right)$, the lower bound of $\langle \hat\Omega  \rangle$ is shown to be  $ |[ \hat\omega_+(\xi),  \hat \omega_-(\xi)]| % \langle \Delta(\hat\omega_+(\xi))\rangle \langle \Delta (\hat\omega_-(\xi))\rangle 
 = \xi^{-1} - \xi$. This minimum of $\hat\Omega $ is achieved by the two-mode-squeezed vacuum (TMSV): %. Its schmidt decomposed form is given by
\begin{eqnarray}
| \psi_\xi \rangle_{AB} &=& \sqrt{1-\xi ^2} \sum_{n= 0}^\infty\xi ^n    |n \rangle  _A |  n \rangle_B \nonumber \\
&= & \sqrt{1-\xi ^2} e^{\xi a^\dagger b^\dagger}|0 \rangle  _A |  0 \rangle_B, \label{defTMSV}
\end{eqnarray}where %  is the number states and the annihiration operators of mode $A$ and $B$, have the relation 
%A we used the relation for the number states of mode $A (B)$, $|n\rangle_{A(B)} $  with the annihilation and creation operators $a$ $a^\dagger $ assoiciated with the 
 the number states are given by %$| n \rangle_{A(B)} $ %:= (\hat a^\dagger )^n | 0 \rangle_A / \sqrt{n !}$ is the Fock basis of the mode $a$
%\begin{eqnarray} | n \rangle_{A } :=  \frac{(a^\dagger )^n}{\sqrt{n!}} |0 \rangle_A, \ \ | n \rangle_{B } := \frac{(b^\dagger )^n}{\sqrt{n!}} |0 \rangle_B$ %\end{eqnarray}
$ | n \rangle_{A } :=   (a^\dagger )^n|0 \rangle_A /{\sqrt{n!}} $ and $| n \rangle_{B } :=  {(b^\dagger )^n}|0 \rangle_B/{\sqrt{n!}}  $ %\end{eqnarray}
associated with the annihilation operators of mode $A$ and $B$, $a := (\hat x_A + i\hat p_A)/\sqrt 2 $ and $b := (\hat x_B + i\hat p_B)/\sqrt 2 $, the vacuum states defined by $a |0 \rangle_A =0$ and $b |0 \rangle_B =0$, and the commutation relations: $[a,a^\dagger ]=1$, $[b,b^\dagger ]=1$, and $[a,b ]= [a,b^\dagger ]=0.$ 
%\begin{eqnarray}[a,a^\dagger ]=1, \  \  [b,b^\dagger ]=1,   \  \    [a,b ]= [a,b^\dagger ]=0. \end{eqnarray}

Using these ladder operators and their commutation relation we can write 
\begin{eqnarray}
%\left\langle 
\hat\Omega (\xi ) 
&=& 2\left(\frac{a^\dagger}{\sqrt \xi }  -  {\sqrt \xi}b   \right)\left(\frac{a}{\sqrt \xi } -  {\sqrt \xi} b^\dagger   \right) +\left(\frac{1}{\xi} -\xi \right)
 \nonumber . % \\
%&=& 2\left(\frac{1}{\xi} -\xi \right)\left(\hat A_\xi ^\dagger \hat A_\xi +\frac{1}{2} \right)
\end{eqnarray}
%where $a := (x_A + ip_A)/\sqrt 2 $ and $b := (x_B + ip_B)/\sqrt 2 $ are the annihiration operators of mode $A$ and $B$, respectively.  %ere we assume $\langle a \rangle= \langle b \rangle =0 $, $0 \le \xi <1 $ for a simplicity.
 Let us define
the new annihilation operator by 
\begin{eqnarray}
 \hat A_\xi :=  \frac{a  -\xi b ^\dagger }{\sqrt{1-\xi^2}}. \label{defA}
 \end{eqnarray}
 This implies $[\hat A_\xi ,\hat A_\xi ^\dagger] =1 $ and $\hat\Omega  (\xi ) $ takes the form of the Hamiltonian of a single harmonic oscillator,
\begin{eqnarray}
\hat\Omega (\xi) \label{s-harm}
%&=& 2\left(\sqrt \xi a - \frac{b^\dagger }{\sqrt \xi}   \right)\left(\sqrt \xi a^\dagger - \frac{b }{\sqrt \xi}   \right) +\left(\frac{1}{\xi} -\xi \right) \nonumber \\
&=&  \left( {\xi}^{-1} -\xi \right)\left(2 \hat A_\xi ^\dagger \hat A_\xi +1 \right).
\end{eqnarray}
From this expression it is clear that $\langle \hat\Omega  (\xi ) \rangle$ is no smaller than $(\xi^{-1} -\xi)$ since $\hat A^\dagger \hat A \ge 0$.
The state that achieves the minimum of $\hat\Omega (\xi)$ is also derived from the condition $\hat A_\xi |0,0;\xi \rangle = 0$. % ground state of this hamiltonian $|0,0;\xi \rangle$ that satisfies the condition $\hat A_\xi |0,0;\xi \rangle = 0$. 
From Eqs (\ref{defTMSV}) and (\ref{defA}) we can verify $\hat A_\xi |\psi_\xi\rangle_{AB} =0$, and thus denote $|0,0;\xi \rangle : = |  \psi_\xi\rangle_{AB}$. The eigen states of the number operator $\hat N_A:=\hat A_\xi^\dagger \hat A_\xi$ is generated by  $(\hat A_\xi ^\dagger  )^{N_A  }|0,0;\xi  \rangle  / \sqrt{N_A!}=: |N_A ,0; \xi  \rangle $ and the eigen value $N_A$ is a non-negative integer. 
The eigen state $ |N_A,0;\xi \rangle$ reduces to $|N_A\rangle_A|0\rangle_B$ in the limit $\xi\to 0$. This suggests that there is another harmonic oscillator whose eigen-state reduces to 
$|0\rangle_A|N_B\rangle_B$ in $\xi\to 0$. We can find another pair of the EPR-like operators, $\hat \omega_\pm (\xi^{-1}) $, which commute with the other ones as $[ \hat \omega_\pm(\xi ), \hat \omega_\pm(\xi^{-1} ) ]=0 $. 
%$ \sqrt \xi \hat p_A + \frac{\hat p_B}{\sqrt \xi } =\hat \omega_- (\xi^{-1}) $. Note that the two of the pairs of the EPR-like operators are commutable each other, $[ \hat \omega_\pm(\xi ), \hat \omega_\pm(\xi^{-1} ) ]=0 $. %, which given by the replacement $ \xi \to \xi^{-1}$replaced with 
Associated with $\hat \omega_{\pm}(\xi^{-1})$ we define another annihilation operator, 
\begin{eqnarray}
\hat B_\xi:=\frac{b  -\xi a^\dagger}{\sqrt{1-\xi ^2}}.\end{eqnarray}
Then, we can write the sum of the total variance of the two pair of the EPR-like operators as 
\begin{eqnarray}
 \hat\Omega (\xi)+\hat\Omega (\xi^{-1})%\nonumber \\  &=&\hat \omega_+^2 (\xi ) +\hat \omega_-^2 (\xi ) + \hat \omega_+^2 (\xi^{-1}) +\hat \omega_-^2 (\xi^{-1}) \nonumber \\
&=&  2\left(\frac{1}{\xi} -\xi \right)\left(\hat A_\xi ^\dagger \hat A_\xi +\hat B_\xi ^\dagger \hat B_\xi + 1 \right). \label{tpepr}\end{eqnarray}
Noting that $\hat B_\xi |0,0;\xi\rangle =0 $ and $[\hat A_\xi,\hat B_\xi]=[\hat A_\xi,\hat B_\xi^\dagger]= 0 $, we obtain all the eigen states of $\hat\Omega (\xi)+\hat\Omega (\xi^{-1})$: %$\hat A_\xi ^\dagger \hat A_\xi +\hat B_\xi ^\dagger \hat B_\xi $: % 
\begin{eqnarray}
|N_A,N_B;\xi \rangle =\frac{ (\hat A_\xi^\dagger)^{N_A} }{\sqrt{N_A!}} \frac{(\hat B_\xi^\dagger)^{N_B}}{\sqrt{N_B!}} |0,0;\xi \rangle . \label{basisi}
\end{eqnarray}
We call this state the  \textit{entangled number state} (ENS) or \textit{two-mode-squeezed number state} \cite{TMSN,Dell06}. 
 While the vacuum state $|0,0;\xi\rangle$ is a Gaussian state, the excited states are NG states. This can be seen by combining the orthonormal relation $\langle N_A',N_B';\xi |N_A,N_B;\xi\rangle =\delta_{N_A,N_A'} \delta_{N_B,N_B'} $ with the fact that a state orthogonal to a Gaussian state is a NG state.  %
 
Note that not only the TMSV but also the states $|0, N_B;\xi\rangle$ with arbitrary excitation of $N_B$ achieve the minimum of $\hat\Omega (\xi)$. % . 
This implies that the minimum is also achieved by the infinite sequence of the NG states  $\{|0,N_B;\xi\rangle | N_B\ge 1 \}$ and these states are entangled.
Formally, the eigen space of $\hat N_A$ specified by $N_A = 0$ is spanned by $\{|0, N_B;\xi\rangle | N_B \ge 0 \}$, and any state belongs to this eigen space achieves the minimum. From the separable criterion of Eq. (\ref{Duan-c}) we can obtain a somewhat strong statement: any state spanned by $\{ |N_A, N_B;\xi\rangle |0 \le N_A < \xi^2/(1-\xi^2), N_B\ge 0  \}$ is entangled.    
 
In order to see that the ENSs are entangled state, we will write $|N_A,N_B;0\rangle$ in the photon number basis  $\{ |n\rangle_A|m \rangle_B \}$. In what follows we assume $N_A\ge N_B$ for a simplicity.
Using the relation $e^{-\xi a^\dagger b^\dagger} b e^{\xi a^\dagger b^\dagger}= b +\xi a^\dagger $ we have $(\hat A_\xi^\dagger)^{N_A} e^{\xi a^\dagger b^\dagger}= (1-\xi^2)^{-N_A/2}[a^\dagger (1-\xi ^2) - \xi b]^{N_A}$. 
By acting this operator on the vacuum, we obtain $|N_A,0 ;\xi\rangle =(\hat A_\xi^\dagger)^{N_A}\sqrt{1-\xi^2} e^{\xi a^\dagger b^\dagger} |0\rangle_A|0\rangle_B /\sqrt{N_A!}= (1-\xi^2)^{(N_A+1)/2} e^{\xi a^\dagger b^\dagger} |N_A\rangle|0\rangle  $.
Similarly, from the relation  $e^{-\xi a^\dagger b^\dagger} a e^{\xi a^\dagger b^\dagger}= a +\xi b^\dagger $ we have $|N_A,N_B ; \xi\rangle   = (1-\xi^2)^{(N_A-N_B+1)/2} e^{\xi a^\dagger b^\dagger} [b^\dagger (1-\xi ^2) - \xi a]^{N_B} |N_A\rangle|0\rangle /\sqrt{N_B!}    $. Then we find the Schmidt decomposed form of the ENS:
\begin{eqnarray}
%|N_A,N_B; \xi\rangle &=& (1-\xi^2)^{\frac{N_A-N_B+1}{2}} \sum_{n=0}^{\infty} \frac{\xi ^n }{n!} \sum_{k=0}^{N_B} \frac{N_B!}{k!(N_B-k)!}(1-\xi^2)^k  ( -\xi  )^{N_B-k}    { { (b^\dagger)^{ n} }{ }}  (a^\dagger)^n   a  ^{N_B-k}  |N_A   \rangle_A |N_B  \rangle_B \nonumber \\
&& |N_A,N_B; \xi\rangle \nonumber \\
&=& (1-\xi^2)^{\frac{N_A-N_B+1}{2}} e^{\xi a^\dagger b^\dagger} \sum_{k=0}^{N_B} \frac{N_B!(1-\xi^2)^k ( -\xi )^{N_B-k}  }{k!(N_B-k)!}  \nonumber \\
 &\times  &\sqrt{\frac{k!}{N_B! }}  \sqrt{\frac{N_A!}{(N_A-N_B+k)! }}  |N_A-N_B+k \rangle_A |k\rangle_B \nonumber \\
%|N_A,N_B; \xi\rangle &=& (1-\xi^2)^{\frac{N_A-N_B+1}{2}} \sum_{n=0}^{\infty} \frac{\xi ^n }{n!} \sum_{k=0}^{N_B} \frac{N_B!}{k!(N_B-k)!} (1-\xi^2)^k  \sqrt{\frac{(k+n) !}{N_B! }} ( -\xi )^{N_B-k} \sqrt{\frac{N_A!}{(N_A-N_B+k+n)! }}  |N_A-N_B+k+n  \rangle_A |k+n \rangle_B \nonumber \\
 &=&  \sum_{m=0}^{\infty }  C_m(N_A,N_B, \xi^2)  |N_A-N_B+m \rangle_A |m \rangle_B %\nonumber \\ &=& (1-\xi^2)^{\frac{N_A-N_B+1}{2}} \sum_{m=0}^{\infty } \sum_{k=0}^{N_B}\left( (1-\xi^2)^k ( -\xi )^{N_B-k} \xi^{m-k}  \frac{\sqrt{N_A! N_B! (N_A-N_B+m)! m!} }{k!(m-k)!(N_A-N_B +k)!(N_B-k)!} \right)  |N_A-N_B+m \rangle_A |m \rangle_B
 \end{eqnarray}
where the Schmidt coefficient $C_m$ is  given by \cite{TMSN} 
\begin{eqnarray}
&& C_m(N_A,N_B, \xi^2) \nonumber\\
 &=& (1-\xi^2)^{\frac{N_A-N_B+1}{2}}  \sum_{k=0}^{\min\{m,N_B\}}\Bigg( (1-\xi^2)^k ( -\xi )^{N_B-k} \nonumber  \\
 & &\times   \xi^{m-k} \frac{\sqrt{N_A! N_B! (N_A-N_B+m)! m!} }{k!(m-k)!(N_A-N_B +k)!(N_B-k)!} \Bigg) .\nonumber \\ \end{eqnarray}
%The normalizationn condition $\sum_{m=0}^{\infty}|C_m|^2 =1 $ is satisfied by the construction.
 Physically, $ \{|C_m|^2 \}$ means the photon number distribution of the system $B$ (or $A$). % is  $\hat n_b : =  b^\dagger b =  m$. 
Actually, %using the inverse relation \begin{eqnarray}a&=&\frac{\hat A_\xi +\xi B_\xi ^\dagger }{\sqrt{1-\xi^2}}, \ \ a ^\dagger=\frac{\hat A_\xi ^\dagger   +\xi B_\xi}{\sqrt{1-\xi^2}} \nonumber\\b&=&\frac{\hat B_\xi +\xi A_\xi ^\dagger }{\sqrt{1-\xi^2}}, \ \ b  ^\dagger = \frac{\hat B_\xi ^\dagger +\xi A_\xi }{\sqrt{1-\xi^2}},\end{eqnarray}
we have the mean  and variance of the number operator $\hat n_b:  =  b^\dagger b  $, %
\begin{eqnarray}
\bar m &: = &\sum_{m=0}^\infty m|C_m |^2 %\nonumber \\
  =   \langle \hat n_b \rangle \nonumber %_{N_A,N_B,\xi}  \nonumber\\
%&=&\frac{1}{\sqrt{1- \xi^2}} \left\langle \hat B^\dagger \hat B + \xi( \hat B^\dagger \hat A+ \hat A^\dagger \hat B)+\xi^2 \hat A  \hat A^\dagger  \right\rangle \nonumber\\
 =  \frac{N_B+ \xi^2 (N_A+1) }{ {1- \xi^2}} \\  %\left(N_B+ \xi (N_A+1)  \right)% \rangle \nonumber\\
\Delta^2 m %:&=& \sum_{m=0}^\infty m^2|C_m |^2 -\bar m ^2\nonumber \\
 &:=& \sum_{m=0}^\infty (m-\bar m)^2 |C_m |^2 \nonumber \\
 & =& \langle \hat n_b^2 \rangle-\langle \hat n_b \rangle^2 %\langle  (b^\dagger   b)^2  \rangle -\langle  b^\dagger   b  \rangle ^2 %\nonumber   \\
%&=&\frac{1}{\sqrt{1- \xi^2}} \left\langle \hat B^\dagger \hat B + \xi( \hat B^\dagger \hat A+ \hat A^\dagger \hat B)+\xi^2 \hat A  \hat A^\dagger  \right\rangle \nonumber\\
  =   \frac{\xi^2 (N_A+ N_B+ 2N_AN_B+1)}{ (1- \xi^2)^2} .  \label{deltam}\nonumber \\%\left(N_B+ \xi (N_A+1)  \right)% \rangle \nonumber\\
\end{eqnarray}
%where $\langle  \hat O \rangle_{N_A,N_B,\xi} := \langle N_A,N_B;\xi  |\hat O|  N_A,N_B;\xi \rangle $
Since the distribution has non-zero variance $\Delta^2 m >0$ whenever $\xi >0 $, any of $C_m$ is less than unity and thus the ENS cannot be a product state. % Therefore all the ENS cannot be a product state. In our case, we have $|C_0|^2 = (1-\xi^2)^{(1+N_A-N_B)}(\xi^2)^{N_B}\frac{N_A! }{(N_A-N_B)! N_B! }<1$ whenever $0<\xi<1$.
 Therefore, all of the ENSs are entangled states.  % The schmidt coefficient with 

The degree of entanglement for the pure states are characterized by the von-Neumann entropy of its marginal state \cite{EOF}. The entanglement corresponds to   
 the Shannon entropy for the distribution of the squared Schmidt coefficients:  %$ \rho_A$: 
%$\textrm{Tr} \hat \rho_A \log_2 \rho_A$ where $\rho_A := \textrm{Tr}_B \hat \rho_{AB} $
\begin{eqnarray}
%&& \textrm{Tr} \hat \rho_A \log_2 \rho_A \nonumber \\ % \hat  a_{n, N }(\xi ):= (1-\xi ^2  )^{\frac{ N +1}{2}} \xi ^{ n} \sqrt\frac{(N+ n)!}{N! +n!}.
E (|N_A,N_B; \xi \rangle )&= &-\sum_{m=0}^\infty |C_m|^2 \log_2 |C_m |^2.
\end{eqnarray}
It reduces to the entanglement of the TMSV when $N_A=N_B=0$ \cite{Gie03}. %, $ E(|0,0;\xi\rangle )=  \frac{1}{1-\xi^2} \log_2 \frac{1}{1-\xi^2} - \frac{\xi^2}{1-\xi^2}  \log_2 \frac{\xi^2}{1-\xi^2}  $.

 \begin{figure}[tb]
  \begin{center}
 \includegraphics[width=8.6cm]{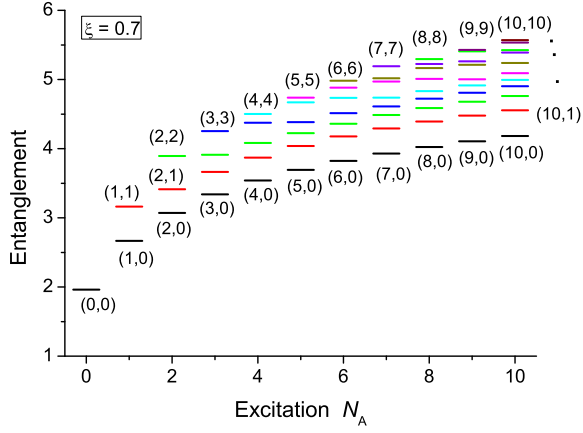}%{G15-EOF10-10.eps}
  \end{center}
  \caption{The entanglement for the entangled number states $|N_A,N_B;\xi\rangle $ with $\xi =0.7$ and $0\le N_A \le  N_B \le 10$. The state is specified by $(N_A,N_B)$.   }
     \label{fig:Fig2.eps}
\end{figure}

\begin{figure}[tb]
  \begin{center}
    \includegraphics[width=8.6cm]{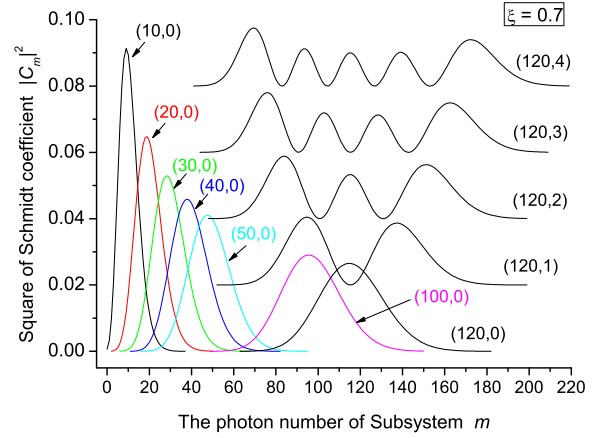}%{Graph-x07.eps}
  \end{center}
  \caption{The distributions of the squared Schmidt coefficient $|C_m|^2$ of the entangled number states $|N_A,N_B;\xi\rangle $ with $\xi=0.7$. The state is specified by $(N_A,N_B)$. %denotes the eigen values and specifies the .
   For $N_B=0 $, % The coefficient of $|N_A,0;\xi\rangle $
  the coefficient  $|C_m|^2$ is a normal wave packet whose mean is $(N_A+1)\xi/(1-\xi^2)$  and variance is $(N_A+1)\xi^2/(1-\xi^2)^2$. An excitation of the $N_B$ mode induces a node on the distribution as is shown in the case of $N_A=120$ and $N_B =0,1,2,3,4$, where $|C_m|^2 + 0.02 N_B$ is displayed instead of $|C_m|^2$.   }%
    \label{fig:Fig1.eps}
\end{figure}

Figure \ref{fig:Fig2.eps} displays $E (|N_A,N_B; \xi \rangle )$ for $\xi=0.7$ and $\{ (N_A,N_B) |0\le N_B \le N_A  \le 10\} $. In order to grasp the entangled property of the ENS, it might be instructive to consider the distribution $ \{|C_m|^2 \}$ for one-mode excitations, e.g., we set $N_B=0$. In these cases, $ \{|C_m|^2 \}$ corresponds to the Negative Binomial (NB) distribution  
\begin{eqnarray}
|C_m(N_A,0,  p)|^2 =  (1-p  )^{ { 1+ N_A} }p ^{ m}  \frac{ (N_A+m)! }{N_A ! m!},
\end{eqnarray}
 and the reduced state of $|N_A,0;\xi \rangle$ is a NB state \cite{NBS}.   
As is shown in FIG. \ref{fig:Fig1.eps}, the NB distribution has a single-peaked waveform whose mean is $\frac{p(N_A+1)}{1-p}$ and variance is $\frac{p(N_A+1)}{(1-p )^2}$. It is also known that the NG distribution approaches to the Poisson distribution $P(k;\lambda)=\lambda^k e^{-\lambda} /k!$ with $\lambda =(N_A+1)(1-p)/p$ for larger $N_A $. The excitation of $N_B$ modulates the normal waveform of the NB distribution and adds the nodes as exemplified in FIG. \ref{fig:Fig1.eps}.  
The parameter $\xi$ determines the typical scale of the distribution [see also Eq.(\ref{deltam})], and we can see that the distribution becomes broader as  $\xi \to 1$.
From these observations on the distribution of $|C_m|^2$ and the fact that the entropy becomes higher if its distribution becomes more uniform, it is likely that the ENS has the following property of entanglement: (i) The entanglement of ENS becomes larger when $N_B$ becomes larger with fixed $N_A$; (ii) The entanglement of ENS becomes larger when $N_A$ becomes larger with fixed $N_B$; and (iii) The entanglement of ENS becomes larger when $\xi$ becomes larger  with fixed $N_A$ and $N_B$. Similar conjectures have also been presented in \cite{Dell06}. %

The ENS is an eigen states of the number operator $\hat A_\xi^\dagger\hat A_\xi$, and the variance $ \langle \Delta ^2 (\hat A_\xi^\dagger\hat A_\xi)\rangle$ is zero. Hence, it is likely that the state who has a small variance of $\hat A_\xi^\dagger\hat A_\xi$ is also entangled. To find out such a relation, let us consider the partial transposition (PT) \cite{NPT} of the variance,  %of the number operator 
$ \langle (\hat A_\xi^\dagger\hat A_\xi)^2  \rangle_{PT} - \langle\hat A_\xi^\dagger\hat A_\xi \rangle_{PT}^2$. Calculating the PT by using the relation \cite{SV1} %$\langle ( a^\dagger)^k a^l ( a^\dagger)^m a^n ( b^\dagger)^o b^p ( b^\dagger)^q b^r   \rangle_{PT}=\langle ( a^\dagger)^k a^l ( a^\dagger)^m a^n \{ ( b^\dagger)^o b^p ( b^\dagger)^q b^r \}^\dagger   \rangle =\langle ( a^\dagger)^k a^l ( a^\dagger)^m a^n \{ ( b^\dagger)^r b^q ( b^\dagger)^p b^o   \rangle $
$\langle  a^{\dagger k} a^l  a^{\dagger m} a^n  b^{\dagger o} b^p  b^{\dagger  q} b^r   \rangle_{PT}=\langle  a^{\dagger k} a^l  a^{\dagger m} a^n  (   b^{\dagger o} b^p  b^{\dagger  q} b^r    )^\dagger   \rangle =\langle  a^{\dagger k} a^l  a^{\dagger m} a^n    b^{\dagger r} b^q  b^{\dagger p} b^o   \rangle $ 
we have the inequality
\begin{eqnarray}
&& (1-\xi^2) ^2 \left( \langle (\hat A_\xi^\dagger\hat A_\xi)^2  \rangle_{PT} - \langle\hat A_\xi^\dagger\hat A_\xi \rangle_{PT}^2 \right) \nonumber \\ \label{pta}%& =&  \langle \Delta^2 [(a^\dagger -\xi b^\dagger  )(a^\dagger -\xi b^\dagger  ) ]  \rangle  - \langle\hat A_\xi^\dagger\hat A_\xi \rangle 
& =&  \langle \Delta^2 [(a^\dagger -\xi b^\dagger  )(a  -\xi b ) ]  \rangle  + \xi^2 
\ge \xi ^2 ,\end{eqnarray} which holds for any valid density operator. The PT of the density operator is also a valid density operator for separable states. From this fact with Eqs. (\ref{s-harm}) and (\ref{pta}) we have a separable criterion: Any separable state must satisfy 
\begin{eqnarray}
\langle \Delta ^2 [\hat \Omega(\xi)  ] \rangle =  \langle (\hat \Omega(\xi) )^2  \rangle - \langle  \hat \Omega(\xi)  \rangle^2  \ge 4.  \label{CRID}\end{eqnarray}
%\langle \Delta ^2 [(\hat A_\xi^\dagger\hat A_\xi)  ] \rangle =  \langle (\hat A_\xi^\dagger\hat A_\xi)^2  \rangle - \langle\hat A_\xi^\dagger\hat A_\xi \rangle^2  \ge \frac{\xi ^2}{ (1-\xi^2) ^2  } \end{eqnarray}
 This inequality is maximally violated by the ENSs, which yield $\langle \Delta ^2 [\hat \Omega(\xi)  ] \rangle =0$, %$\langle \Delta ^2 [(\hat A_\xi^\dagger\hat A_\xi)  ]\rangle =0 $.
% We have the same criterion associated with $\hat B_\xi^\dagger\hat B_\xi$,
  and thus ENSs are negative PT entangled states. For a given $N_A$, any state spanned by $\{|N_A,N_B;\xi \rangle | N_B\ge 0 \}$ also maximally violates the inequality (\ref{CRID}). Interestingly, the two separable criteria of Eqs. (\ref{Duan-c}) and (\ref{CRID}) are concerned with quite different aspects of the same operator $\hat \Omega (\xi)$ of Eq. (\ref{noise-op}). While Criterion (\ref{Duan-c}) is of relevant to the Gaussian states \cite{Duan-Simon} our Criterion (\ref{CRID}) is, at least, capable of detecting the eigen states of Eq. (\ref{basisi}), which form a family of the complete orthonormal set specified by the parameter $\xi $.

The ENS is an eigen state of the quadratic Hamiltonian of Eq.(\ref{tpepr}), and thus the lower exited states as well as TMSV will appear in such physical systems. By noting that the vacuum $|0\rangle_A |0 \rangle_B $ and TMSV $ |0,0;\xi\rangle$ are connected with the unitary operator of the parametric amplification $U= e^{  (a^\dagger b^\dagger - ab)r }$ as $|0,0;\xi\rangle = U |0\rangle_A |0 \rangle_B$ with $r= \tanh^{ -1} \xi$ and from the relation $ \hat A_\xi^\dagger  |0,0;\xi\rangle = U a^\dagger  U^\dagger U |0\rangle_A |0 \rangle_B = U a^\dagger |0\rangle |0 \rangle$, the ENS will appear as the output of the parametric amplification when the input state is a number state \cite{Dell06,TMSN,NBS}. The operator $\hat A^\dagger \propto a^\dagger +\xi b $ implies the process where the addition of a photon on the mode $A$ and the subtraction of a photon in the mode $B$ are done coherently. Hence ENS might also be realized by performing such process on the TMSV.

%Simultaneous  eigenstates of $\hat A_\xi$ and $\hat B_\xi$
Given the ``number-state'' basis we may define the ``coherent state'' or ``squeezed states''. We define the ``coherent state'' as the simultaneous eigen state of the annihilation operators: $\hat A_\xi|\alpha,\beta ; \xi\rangle  = \alpha  |\alpha,\beta ; \xi\rangle $ and $\hat B_\xi|\alpha,\beta ; \xi\rangle  = \beta  |\alpha,\beta ;\xi\rangle $ with complex amplitudes $\alpha$ and $\beta$.
It can be represented in the basis of the ENSs as
\begin{eqnarray}
|\alpha ,\beta; \xi\rangle &= &e^{- \frac{|\alpha |^2}{2} } e^{- \frac{|\beta |^2}{2} }  \sum_{n=0}^{\infty}\sum_{m=0}^{\infty} \frac{\alpha ^n}{\sqrt{n!}}\frac{\beta^m}{\sqrt{m!}}|n,m;\xi\rangle \nonumber \\
&=& \hat D_{A_\xi}(\alpha ) \hat D_{B_\xi}(\beta )|0,0;\xi\rangle, % \nonumber \\
\end{eqnarray}
where we used the relation $e^{\alpha a ^\dagger -\alpha ^* a }= e^{-|\alpha |^2 /2 }e^{\alpha  a ^\dagger }e^{-\alpha ^* a }$ and introduced the displacement operator %is defined by
\begin{eqnarray}
\hat D_{A_\xi}(\alpha ):=  e^{\alpha A_\xi^\dagger - \alpha^* A_\xi }.\end{eqnarray} % $ e^{\beta B_\xi^\dagger - \beta ^ * B_\xi }$
Since %each of $A$ $A$  $B$ and $B$ is a linear conbination of $a$ $a$ $b$ $b$%the annihilation operator of the original modes $a$ $a$ $D_{A_\xi}(\alpha )=D_{a }\left(  {\alpha}/{\sqrt{1-\xi^2}} \right )D_{b }\left(  {\xi\alpha^*}/ {\sqrt{1-\xi^2}}   \right ) $ and $D_{B_\xi}(\beta)=D_{b }\left(  {\beta}/{\sqrt{1-\xi^2}} \right )D_{a }\left(  {\xi\beta^*}/ {\sqrt{1-\xi^2}}   \right )$  
the displacement operator is written by the product of the local displacement operators %.e.g. $D_{A_\xi}(\alpha )=D_{a }\left(  {\alpha}/{\sqrt{1-\xi^2}} \right )D_{b }\left(  {\xi\alpha^*}/ {\sqrt{1-\xi^2}}   \right ) $. 
 the ``coherent states'' can be converted with each other by the local unitary operation and we have 
\begin{eqnarray}
 |\alpha ,\beta ;\xi\rangle  =\hat  D_{a }\left(  \frac{\alpha+\xi \beta^*}{\sqrt{1-\xi^2}} \right )\hat D_{b }\left(  \frac{\beta + \xi\alpha^*}{\sqrt{1-\xi^2}}   \right )|0,0;\xi\rangle . \nonumber% \\ 
\end{eqnarray}  Hence, the ``coherent states'' are equivalent to the TMSV under the local unitary transformation, and the amount of entanglement is equal to that of the TMSV. The displaced TMSV states are also called as the two-mode-squeezed coherent states. % 
 Note that we have the product of the coherent states 
 $|\alpha ,\beta; \xi\rangle  \to |\alpha\rangle _A| \beta \rangle_B $ in the limit $\xi\to 0$. Note also that we can use the expansion by the over complete relation with the equally entangled Gaussian pure states $\int|\alpha ,\beta; \xi \rangle\langle \alpha ,\beta; \xi |d^2\alpha d^2 \beta /\pi^2 =\openone_A\otimes \openone_B $ whereas we have the expansion by the complete orthonormal basis of the ENSs each of which has typically different amount of entanglement $ \sum_{N_A,N_B} |N_A ,N_B; \xi \rangle\langle N_A,N_B; \xi | =\openone_A\otimes \openone_B $. These complementary roles played by the number states and coherent states are also thought to be useful in the theory of entanglement. %

In conclusion we have introduced the entangled number states as the eigen states of a quadratic Hamiltonian. The ground state corresponds to the two-mode-squeezed vacuum and belongs to the Gaussian states. The excited states are Non-Gaussian entangled states. The Schmidt coefficients and the entanglement were calculated. It was conjectured that the higher number state has larger amount of the entanglement. We have also derived a separable criterion from the partial transposition of the variance of the quadratic form. This separable inequality is maximally violated by the entangled number states and these states are shown to have negative partial transposition.   
With their mathematically tractability and well-defined properties due to the eigen states of the harmonic oscillators,  the entangled number states could form a useful testing ground for entanglement theories concerning Non-Gaussian states, and the characterization of the entangled number states will be definitive starting point for understanding the class of Non-Gaussian entangled states.

%\begin{theacknowledgments}
The author would like to thank  Fuyuhiko Tanaka for helpful discussions. %This work was supported by the Grant-in-Aid for the Global COE Program ``The Next Generation of Physics, Spun from Universality and Emergence'' from the Ministry of Education, Culture, Sports, Science and Technology (MEXT) of Japan. 
R.N. acknowledges support by JSPS.
% R.N. is supported by JSPS Research Fellowships for Young Scientists. 
%\end{theacknowledgments}

\end{document}